# Ultrafast and low-energy switching in voltage-controlled elliptical pMTJ

*Jiefang Deng[1,2], Gengchiau Liang[1,2,\*], Gaurav Gupta[3,†]*

[1]Department of Electrical and Computer Engineering, National University of Singapore, Singapore 117576

[2]Centre for Advanced 2D Materials and Graphene Research Centre, National University of Singapore, Singapore 117546

[3]Spin Devices, Delhi 110006, India

Switching magnetization in a perpendicular magnetic tunnel junction (pMTJ) via voltage controlled magnetic anisotropy (VCMA) has shown the potential to markedly reduce the switching energy. However, the requirement of an external magnetic field poses a critical bottleneck for its practical applications. In this work, we propose an elliptical-shaped pMTJ to eliminate the requirement of providing an external field by an additional circuit. We demonstrate that a 10 nm thick in-plane magnetized bias layer (BL) separated by a metallic spacer of 3 nm from the free layer (FL) can be engineered within the MTJ stack to provide the 50 mT bias magnetic field for switching. By conducting macrospin simulation, we find that a fast switching in 0.38 ns with energy consumption as low as 0.3 fJ at a voltage of 1.6 V can be achieved. Furthermore, we study the phase diagram of switching probability, showing that a pulse duration margin of 0.15 ns is obtained and a low-voltage operation (~ 1 V) is favored. Finally, the MTJ scalability is considered, and it is found that scaling-down may not be appealing in terms of both the energy consumption and the switching time for the precession based VCMA switching.

**I. INTRODUCTION**

Spin transfer torque (STT) based magnetic random access memory (MRAM) [1, 2] because of its non-volatility, high access speed and CMOS (complementary metal oxide semiconductor) compatibility [3] has matured into one of a leading candidate in recent years [4] to fill the memory gap in the extant memory hierarchy. A bit-cell of a STT-RAM comprises of a magnetic tunnel junction (MTJ) which has at least one pinned-layer (PL) and at least one free-layer (FL) with their magnetization in either parallel (P) or anti-parallel (AP) state with respect to (w.r.t.) each other, which corresponds to logic "1" or "0". In STT-RAM, a free-layer is written by passing a current with density larger than critical current density through the MTJ. If the current is flowing from FL towards PL, a spin-flux with vector parallel to the magnetization of the PL ($\mathbf{M_{PL}}$) acts on the FL to align the magnetization of the FL ($\mathbf{M_{FL}}$) with $\mathbf{M_{PL}}$, whereas if the current direction is reversed a reverse spin-flux with vector anti-parallel to the $\mathbf{M_{PL}}$ acts on the FL to orient it in the AP state. This current based magnetization switching, however, requires a large current density which ranges from $5\times10^{10}$ A-m$^{-2}$ for 35 ns to $1\times10^{11}$ A-m$^{-2}$ for 2 ns write-time [1, 2] to generate enough spins to toggle all the magnetic moments of the FL. Inevitably, a large current results in considerable Joule heating in the MTJ. This results in self-heating [5-8] induced degradation of the MTJ characteristics, e.g. the spin-polarization of the spin-flux degrades thereby degrading the STT efficiency at the higher temperatures. In addition, electromigration [9, 10] becomes prominent because of the large current densities and the dielectric may also break [11, 12] at the voltages required to sustain the required current densities. Furthermore, to provide a large enough current a bulky access transistor, i.e. with a large channel width-to-length ratio for a bulk-MOSFET or with a large number of fins for a FinFET (Fin Field Effect Transistor), is required. This implies that STT-RAM suffers from high energy consumption, reliability issues and a huge cell area [13, 14].

To reduce the operating current which would subsequently reduce both the energy consumption in the device and the size of the driving transistor, a voltage control of magnetic anisotropy (VCMA) [15] has been promulgated as an alternative to the STT [16, 17]. Since the VCMA based device relies on a voltage rather than a current to write, the current in this case can be greatly reduced by designing a MTJ of a large enough resistance. Furthermore, the VCMA based precessional switching enables the toggling of the FL in sub-ns time regime [15, 18]. Therefore, the energy consumption can be substantially reduced by markedly reducing both the power-dissipation and the dissipation-time. The demonstrations [18-20] hitherto have been on the individual circular pMTJs. These demonstrations require an external in-plane magnetic field to enable the switching. This, however, is not a viable solution for the integrated MRAM w.r.t. both the provision of an external field source and the field uniformity [21, 22]. Consequently, the requirement of an external bias field poses a critical bottleneck for realizing a practical VCMA memory.

In this work, therefore, we propose an elliptical pMTJ to eliminate the requirement of an external magnetic field source. Because of the elliptical structure, via shape anisotropy, an in-plane magnetized bias-layer (BL) separated from the FL by a metallic spacer can be directly engineered within the MTJ stack [23]. This in-plane BL, hence, provides a sufficient bias field for the VCMA based precessional switching. We comprehensively appraise the effects of the electronic, magnetic and physical design constraints on the FL switching dynamics to expound the device physics and an optimal operation window in the proposed design. Our results show that the required bias magnetic field can be contrived within the MTJ stack by an in-plane magnetized BL. For instance, a 10 nm thick BL separated by a 3 nm thick metallic spacer can provide an in-plane exchange field of 50 mT to bias the FL. Furthermore, we show that the FL can toggle in just 0.38 ns consuming only 0.3 fJ at a voltage of 1.6 V across the MTJ, which is attractive for the memory applications. Our results also indicate that the pMTJ driven by the precession based VCMA favors a low-voltage operation (~ 1 V) with a sufficient margin (0.15 ns) for the applied voltage pulse duration.

This paper is organized as follows. Section II introduces the theoretical method for the simulations. Section III elaborates the operation mechanism for this precession based VCMA switching, followed by the discussion and analysis of different parameters affecting the device properties including a voltage across the MTJ, MgO thickness $t_{MgO}$, a bias magnetic field and the scalability of the MTJ cross-section. This is followed in Sec. IV by the summary and conclusion of our study.

**II. METHODS**

Figure 1(a) shows a schematic of an elliptical MTJ with perpendicular magnetic anisotropy (PMA), with major-axis (minor-axis) along x-axis (y-axis). A 1.2 nm thick $Co_{20}Fe_{60}B_{20}$ [24] PL and a 1.7 nm thick $Co_{20}Fe_{60}B_{20}$ FL sandwich a MgO insulator of thickness $t_{MgO}$. In this work, $t_{MgO}$ ranges from 1.2 nm to 2.8 nm, while the MTJ cross-section ranges from 150×50 $nm^2$ to 114×38 $nm^2$ with a fixed aspect-ratio (AR) of 3. Macrospin simulation, which has been shown to be valid in purview of dimensions considered in this work [15, 18, 25, 26], is developed to investigate the magnetization dynamics. The dynamics is described by the Landau-Lifshitz-Gilbert (LLG) equation [27-29] as,

$$\frac{\partial \mathbf{m}}{\partial t} = -\gamma \, \mathbf{m} \times \mu_0 \mathbf{H}_{Eff} + \alpha \, \mathbf{m} \times \frac{\partial \mathbf{m}}{\partial t} + \Gamma_{MTJ\_DL} \, \mathbf{m} \times \mathbf{p}_{PL} \times \mathbf{m} + \Gamma_{SV} \, \mathbf{m} \times \mathbf{p}_{BL} \times \mathbf{m} + \Gamma_{MTJ\_FL} \, \mathbf{m} \times \mathbf{p}_{PL}, \quad (1)$$

where $\gamma$ is the gyromagnetic ratio, $\mu_0$ is the vacuum permeability, $\alpha = 0.01$ [28] is the Gilbert damping coefficient for $Co_{20}Fe_{60}B_{20}$, and $\mathbf{p}_{PL}$ ($\mathbf{p}_{BL}$) is a unit-vector anti-parallel (parallel) to the magnetization of the PL (BL). The magnetization unit-vector of the FL $\mathbf{m}$ is [$m_x$ $m_y$ $m_z$], which is [0 0 ±1] in stable states, with $m_x$, $m_y$ and $m_z$ being the projections on the respective axis denoted by the subscripts. $\mathbf{H}_{Eff}$ is the effective magnetic field experienced by the FL. It is the vector sum of a uniaxial anisotropy field $\mathbf{H_K}$, a demagnetizing field $\mathbf{H_D}$, a thermal fluctuation field $\mathbf{H_{Therm}}$, and an external bias field $\mathbf{H_{Bias}}$, which are expressed as,

$$\mathbf{H_K} = \frac{2\,K_U}{\mu_0\,M_S}[0, 0, m_z], \quad K_U = K_{U\_Bulk} + \frac{K_{I0} - \xi \cdot E_z}{t_{FL}}, \quad (2)$$

$$\mathbf{H_D} = -M_S\left[N_x m_x, N_y m_y, N_z m_z\right], \quad (3)$$

$$\mathbf{H_{Therm}} = \sqrt{\frac{2\,\alpha\,K_B\,T}{(1+\alpha^2)\,\gamma\,M_S\,V(\Delta t)}}\,\frac{1}{\mu_0}[G^x_{(0,1)}, G^y_{(0,1)}, G^z_{(0,1)}], \quad (4)$$

where $K_U$ is the anisotropy energy density with contributions from the bulk anisotropy, $K_{U\_Bulk}$, and the interfacial anisotropy $K_I$. The latter is computed as $K_{I0} - \xi \cdot E_z$. As implied from eq. (2), the interfacial anisotropy is assumed to be linearly modified by the perpendicular component $E_z$ of the electric-field $\mathbf{E}$ at the MgO-FL interface, at a rate determined by the VCMA coefficient $\xi$ [24, 30], where eq. (2) assumes a positive value of $E_z$ for the field direction shown in Fig. 1(a). For the voltage applied across the MTJ $V_{MTJ}$, the magnitude of $E_z$ is assumed to be $V_{MTJ}/t_{MgO}$, assuming an entire potential drop across the insulator [18, 20]. The CoFeB and the CoFeB/MgO interface parameters like saturated magnetization $M_S = 1.257 \times 10^6$ A-m$^{-1}$, $K_{U\_Bulk} = 2.245 \times 10^5$ J-m$^{-3}$, $K_{I0} = 1.286 \times 10^{-3}$ J-m$^{-2}$ [24] and $\xi = 50$ fJ-V$^{-1}$-m$^{-1}$ [31] are the empirical parameters from the experimental papers. The value of $\xi$ at the CoFeB/MgO interface in literature is in the range of 20-100 fJ-V$^{-1}$-m$^{-1}$ [31-34]. More efficient VCMA effect, i.e. a large $\xi$, would result in even better performance of the proposed device than that predicted in this work.

In eq. (3), $N_x$, $N_y$ and $N_z$ are the demagnetizing factors along x, y and z direction, which are determined by the shape and the size of the magnet (shape anisotropy) [35, 36]. The dipole field from the PL has been neglected assuming that this dipole field can be cancelled out by synthetic ferrimagnetic reference layers [37]. Thermal fluctuation is described by eq. (4), where $K_B$, $V$ and $\Delta t$ are the Boltzmann constant, the FL volume and the calculation time step of 5 ps, respectively. The stochastic partial differential equation (SPDE) described by eq. (1) is integrated via fourth order Runge-Kutta method [38-40]. An alternate method for solving this SPDE is first-order Euler method [41, 42], and the trade-offs between the methods are discussed in Refs. [38, 43] for a more interested reader. The device is assumed to operate at the room temperature (T = 300 K), and the self-heating effects due to Joule heating have been ignored because the VCMA devices, including the one discussed in this work with data shown in the next section, operate at much lower current densities than the traditional STT devices. $G_{(0, 1)}$ with superscript along the respective axis are the independent random numbers computed at every time-step and each has a Gaussian distribution with zero mean and unit standard deviation [44]. **$H_{Bias}$** (c.f. Fig. 1(a)) is provided by an in-plane magnetized BL (Co/Pt multilayers) via its dipole field, which is calculated by micromagnetics simulation using MuMax3 simulator [45]. The simulation cell size is chosen to be 1 nm along all three dimensions. Since **$H_{Bias}$** is only along the x-axis in this study, subsequently, it is represented by a scalar quantity $H_x$. The damping-like torque (DLT) with a linear dependence on $V_{MTJ}$, and the field-like torque (FLT) with a quadratic dependence on $V_{MTJ}$ are respectively obtained as,

$$\Gamma_{MTJ\_DLT(SV)} = \frac{\hbar}{2e} \frac{\gamma \, \eta_{MTJ(SV)}}{M_S V \, R_{MTJ}} V_{MTJ} \, , \quad \Gamma_{MTJ\_FLT} = \nu \frac{\hbar}{2e} \frac{\gamma \, \eta_{MTJ}}{M_S V \, R_{MTJ}} V_{MTJ}^2 \, , \qquad (5)$$

where $\hbar$ is the reduced Planck constant, $e$ is the charge of an electron, $R_{MTJ}$ is the resistance of the MTJ, $\eta_{MTJ(SV)}$ is the STT efficiency, and $\nu = 2.97/7.82$ $V^{-1}$ is the ratio between the two torques extracted from Ref. [46] (see Table 2.1 and 2.2 of thesis from Kerstin Bernert [47] for the range of values in literature). Analysis of $R_{MTJ}$ includes the voltage dependence of the tunneling magnetoresistance (TMR) and the dynamic angle θ between the FL and PL as,

$$R_{MTJ} = R_P + \frac{R_{AP0} - R_P}{1 + \frac{V_{MTJ}^2}{V_{Half}^2}} \left( \frac{1 - \cos(\theta)}{2} \right), \qquad (6)$$

where $V_{Half} = 0.4$ V, extracted for CoFeB from Ref. [20, 48], is the voltage across a MTJ at which TMR becomes half of its value at zero-bias i.e. $TMR_0/2$. $R_P$ is the MTJ resistance when both magnets are exactly parallel (P) to each other and assumed to remain invariant to $V_{MTJ}$ for all practical purpose [46, 49], while $R_{AP0}$ is the MTJ resistance when both magnets are exactly anti-parallel (AP) to each other at a zero-bias. Since in the recent years Slonczewski expression [27] for spin-torque efficiency has been extended to account for multiple reflections of the spin-flux in the spin valves [50-53], the STT effect by the BL in this study is based on the multi-reflection model. Hence, the STT efficiencies for the PL-MgO-FL MTJ ($\eta_{MTJ}$) and the FL-Metal-BL spin valve ($\eta_{SV}$) are computed as [52, 54],

$$\eta_{MTJ} = \frac{P_1/2}{1+P_1^2 \cos(\theta)}, \text{where } P_1 = \sqrt{\frac{TMR_0/2}{1+TMR_0/2}},$$
$$\eta_{SV} = \frac{P_2 - P_2 \varsigma \cos(\theta)}{1 - \varsigma^2 \cos^2(\theta)}, \text{where } \varsigma = 1 - 2\varepsilon + 2\varepsilon^2 \text{ and } \varepsilon = \frac{1-P_2}{2}, \quad (7)$$

where $P_1$ is obtained from Julliere's formula for equal polarization of FL and PL [55] and $P_2 = 0.35$ [56].

## III. RESULTS

### A. Operation Principle

Since in a VCMA based MTJ, the interfacial anisotropy energy can be tuned by an applied voltage, the competition between the uniaxial anisotropy and the demagnetizing field, which determines the easy axis direction, can be controlled by the voltage. Specifically, considering the bulk and the interface PMA with the demagnetization simultaneously, the net anisotropy energy density becomes,

$$K_{U\_Eff} = K_U - \frac{1}{2} \mu_0 M_S^2 N_Z. \quad (8)$$

When $K_{U\_Eff} > 0$, FL has a perpendicular magnetization i.e. it has the easy-axis along the z-axis. If $K_{U\_Eff} < 0$, the magnet has the easy-axis along the major-axis of the elliptical FL i.e. the x-axis in this work. Among the parameters in eq. (2) and (8), $t_{FL}$ and $N_Z$ depend on the physical dimensions of the magnet and remain fixed once the MTJ is fabricated in accordance with the design specifications, while $E_Z$ can be dynamically modified by controlling $V_{MTJ}$. For the precession based VCMA switching of the pMTJ devices [18, 20], these physical and electrical controls are designed such that in the absence of a $V_{MTJ}$, $K_U$ which equals $K_{U\_Bulk} + K_{I0}/t_{FL}$ is large enough for $K_{U\_Eff}$ to be positive. Howbeit, the $V_{MTJ}$ is designed to be large enough to have a sufficient $E_Z$ that can render $K_{U\_Eff}$ to a computationally negative value. Physically, this implies that the FL destabilizes along the z-axis and its easy-axis now aligns along the x-axis, thereby forcing **m** of the FL to tend towards a new stable state governed by the **H**$_{Eff}$ and the STT. Conversely, if a negative voltage pulse is applied, as evident from eq. (2) and (8), $K_{U\_Eff}$ becomes more strongly positive because the interface anisotropy is enhanced. Categorically, this has been suggested as a scheme to read MTJ with an increased reliability [57]. The critical values, at which $K_{U\_Eff}$ is zero and thus the orientation of the easy-axis changes, are defined as follows in this study. The critical thickness of FL, in the absence of $E_Z$, is symbolized by $t_{FLC}$, while $V_C$ and $E_{ZC}$, respectively, symbolize the critical values of $V_{MTJ}$ and $E_Z$ for a given $t_{FL}$. Furthermore, the applied $V_{MTJ}$ should not exceed the dielectric breakdown field $E_{Break}$ for MgO insulator [11] which is slightly over 2 V-nm$^{-1}$. Hence, the device is designed such that in absence of a $V_{MTJ}$, $K_{U\_Eff}$ is positive, while it becomes negative for a $V_{MTJ}$ pulse (c.f. Fig. 1(b)) where $E_{ZC} < V_0/t_{MgO} < E_{Break}$. The design is furthermore constrained by a maximum permissible voltage in the system which is not discussed in this study because it is subjective to the targeted application and the desired stability factor Δ of the FL. Δ is computed *in the absence of $V_{MTJ}$* and described as [58],

$$\Delta = \Delta_0 \left(1 - \frac{H_x}{H_K^{Eff}}\right)^2, \text{ where } \Delta_0 = \frac{K_{U\_Eff} V}{K_B T} \text{ and } H_K^{Eff} = \frac{2 K_{U\_Eff}}{\mu_0 M_S}. \quad (9)$$

If the FL is permanently biased as in this work, the stability can be reduced quadratically as a result of the bias field, whereas if $H_x$ is applied only during the write operation [59], $\Delta$ can be substantially increased to equal $\Delta_0$. The scheme suggested in Ref. [59] depends on the Oersted field generated around the current carrying wire in the adjacent cells. Consequently, for a substantial $H_x$ firstly it becomes power intensive and secondly this acts as a stray field and disturbs other bit-cells thereby limiting the memory density. Therefore, an alternative scheme of increasing $\Delta$ without compromising with $V_{MTJ}$ or reducing $H_x$ would be an important future direction.

For a precession based VCMA switching of the pMTJ devices, a $V_{MTJ}$ as a trapezoidal pulse of duration $t_{Pulse}$ is applied to toggle the FL, as shown in Fig. 1(b). A finite rise and fall time, $t_{Rise}$ and $t_{Fall}$, respectively, of 50 ps is assumed to consider a non-ideal input. A full-scale voltage $V_0$, which exceeds $V_C$, is applied for time $t_{High}$ duration to temporarily change the easy-axis from z- to x-axis. A non-zero pulse is then applied for duration $t_{ON} = t_{Rise} + t_{High} + t_{Fall}$. This induces the precession of **m** around the shifted **H**$_{Eff}$. A sufficiently large $H_x$ allows the FL $m_z$ to swing from +1 to −1 and vice versa as illustrated in Fig. 2 (a-c) for the FL of thickness $t_{FL}$ 1.7 nm, cross-section 150×50 nm$^2$, $t_{MgO}$ of 2 nm, $V_0$ of 1.6 V and $\mu_0 H_x$ of 50 mT. As shown in Fig. 2 (a-c), by designing $t_{ON}$ to be odd or even multiples of the half-precession period ($t_{Half}$ = 0.35 ns for the shown cases), the $m_z$ toggles or comes back to the original state, respectively. A long enough time of $t_{OFF}$ ensures that the **m** relaxes to the z-axis. Consequently, the final state of **m** strongly depends on $t_{ON}$ since it determines if **m** is above or below the x-y plane when the $V_{MTJ}$ goes to zero and the easy-axis is switched back along the z-axis. This implies that pulse duration should be controlled in a certain range to have a deterministic switching. Notably, if $t_{High}$ is comparable with or larger than the relaxation time so that **m** would damp to align and lock with the $H_{Eff}$ or if the rise and fall time are comparable with or larger than $t_{Half}$, the switching would become extremely sensitive to the transition times i.e. $t_{Rise}$ and/or $t_{Fall}$. The study of the effect of $t_{Rise}$ and $t_{Fall}$ is, however, not considered in the current work because a voltage pulse with 50 ps rise/fall time can be generated with a jitter at sub 10 picosecond to femtosecond scale [60, 61], which in fact is at least two orders smaller than the $t_{Half}$. The effect of $t_{Rise}$ and $t_{Fall}$ is thus assumed to be negligible compared with other parameters in this study.

Pessimistically, the deterministic switching based on the precise control of the precession cycles requires a bias magnetic field. As mentioned in eq. (1), the $H_{Eff}$ experienced by the FL includes $H_K$, $H_D$ and $H_{Therm}$. Among them, the $H_K$ only has a z-component field while the $H_D$ and the $H_{Therm}$ can be non-zero along the x-axis. However, the small $N_x$ due to the FL dimensions in eq. (3) and the thermal fluctuation as a perturbation are unable to provide a large enough x-component field to support a full-scale precession around the x-axis. As a result, the application of a bias field $H_x$ is seemingly indispensable. To simplify the design for generating the $H_x$, an elliptical pMTJ is used, which allows an in-plane magnetized BL separated by a metal spacer to be fabricated within the MTJ stack (c.f. Fig. 1(a)). The dipole field provided by the BL is shown in Fig. 2(d). It shows that, typically as expected, a thinner metal spacer results in a stronger bias field because a dipole field strengthens as the distance from the magnet decreases. Intuitively, thickening the BL, while ensuring that a single domain is maintained, would have more magnetic moments along the x-axis, which then would lead to a larger bias field for the same $t_M$. Hence, the $t_M$ and the $t_{BL}$ can be custom designed to obtain the required $H_x$. In this work, the $\mu_0 H_x$ applied on the MTJ is approximately 50 mT. As shown in Fig. 2 (d), this large $H_x$ can be provided via a 10 nm thick BL with the metal spacer of $t_M$ 3 nm, implying that an elliptical MTJ stack can function without an additional external system to provide the bias field.

Furthermore, in Fig. 2, the $t_{MgO}$ is thick enough (2 nm) for the chosen $V_0$ to render the STT ineffective. In case the MgO is thin and/or $V_0$ is large enough so that the STT is strong, STT would also play an important role in the FL dynamics as evident from eq. (1). For the configuration in Fig. 1(a), the electrons flow from FL to PL and thus a reflected spin-flux oriented anti-parallel to $\mathbf{M_{PL}}$ acts on the FL, i.e. along the −z-axis in this case and thus STT effect attempts to damp $\mathbf{m}$ towards the −z-axis. This has following important implications on the switching. Pure VCMA effect would be symmetric w.r.t. switching from P-to-AP or vice versa state of the MTJ, but the STT would introduce an asymmetry in the dynamics. Since, [0 0 −1] spins act on the FL, it becomes easier (faster, reliable and more energy efficient) to toggle into or retain $m_z = -1$ state, but it becomes difficult to toggle into or retain $m_z = +1$ state. Therefore, if the PL had been pinned along the z-axis then for the MTJ in presence of a non-negligible STT effect, switching into an AP state would be easier than into a P state.

## B. $V_{MTJ}$ Dependence

First we investigate the electronic control of the device which allows designer to easily and dynamically tune the performance post-fabrication of the device. To probe the effects of $V_0$ on the precession based VCMA switching in an elliptical pMTJ, a representative FL of size 150×50×1.7 nm$^3$ is chosen. A $t_{MgO}$ of 2 nm, resistance-area (RA) product of 1820 Ω-µm$^2$, and a zero voltage-bias tunnel magnetoresistance ratio (TMR$_0$) of 144% is picked from Ref. [20]. A 50 mT bias field along x-axis $H_x$ is applied to assist the switching. This in-plane field, however, reduces the FL thermal stability from 138 to 28 as calculated from eq. (9). This greatly shrunk thermal stability limits the applicability of the proposed elliptical pMTJ in the storage class memories (SCMs) [62, 63] and the enterprise storage [64] which need a retention time on the scale of months to 10 years or more [63-66]. However, because of the ultra-low power sub-ns writing in this design, and the stability which suffices the requirements for an embedded non-volatile cache memory [57, 67, 68], the promulgated precessional VCMA based MRAM may be a promising replacement for a power-intensive volatile static random-access memory (SRAM) in the cache. The detailed appraisal of the performance metrics for a VCMA based MRAM [57] is beyond the purview of our current work and we focus on the device physics of the proposal. Furthermore, this stability also suffices for the MTJ based non-volatile logic (NVL) [69].

The phase diagrams of the switching probability for switching from P-to-AP (P10) and from AP-to-P (P01) are shown in Fig. 3(a) and (b), respectively. For sweeping $t_{ON}$, $t_{High}$ is swept while $t_{Rise}$ and $t_{Fall}$ are held constant at 50 ps each. Each colour point is determined by simulating the device 100 times under the identical conditions while considering thermal fluctuation. Red regions (operation windows), which are directly related to the precession period, signify a deterministic toggling, i.e. 100% probability of switching, while the dark-blue regions denote an unaltered **m** state i.e. 100% probability that the original magnetic-state is retained. For a given $V_0$, the probability oscillates between 0 (the dark-blue regions) and 1 (the red regions) with $t_{ON}$ because for the odd and even multiples of $t_{Half}$, the FL toggles and gets restored to the original state, respectively. However, at small $V_0$ and large $t_{ON}$, the switching is nondeterministic and probability is approximately 50%, as also observed experimentally in Ref. [15]. This is because at low $V_0$ and large $t_{ON}$, the VCMA effect is relatively weak and the $t_{High}$ is comparable with the relaxation time, thus failing to keep the precession about the x-axis for a long time and resulting in an uncertain final state in the presence of the thermal fluctuation. Comparing Fig. 3(a) with 3(b) shows that there is no significant difference in the phase diagram for P10 and P01, which implies that there is symmetry in the switching from P-to-AP and AP-to-P state, indicating that the STT effect is negligible. Next, the black curve with the bars indicates deterministic switching without thermal fluctuation to explicitly illustrate the effect of the thermal fluctuations for the fastest $t_{ON}$ scenario. This $t_{ON}$ is chosen to equal $t_{Half}$. As shown in Fig. 3(a) and (b), the operation window shrinks, as expected, when the thermal fluctuation is considered, implying that for a precession based VCMA switching the thermal fluctuation perturbs the deterministic toggling. Moreover, the operation window expands as the $V_0$ diminishes. This is because the reduced $V_0$ weakens the VCMA effect, resulting in a stronger interfacial anisotropy field along the z-axis as evident from eq. (2). This tends to increase the magnitude of $m_z$ and reduce $m_x$ thus subduing shape anisotropy field along the x-axis, resulting in the reduction of the x-component of **H**$_{Eff}$. The **m** still precesses around the x-axis, although its trajectory is now not totally symmetrical about the x-y plane. Because the precessional period is inversely proportional to the magnetic field along the precession-axis, which is nearly along the x-axis, with a reduced field along the x-axis the precessional period increases. As a result, the operation window, which is closely related to the precession period, increases as $V_0$ decreases. A larger operation window implies more tolerance in variation for $t_{High}$, and hence, a more reliable write-operation, indicating that the low-voltage operation is achievable for this studied device.

The $V_0$ dependence of the current density $J$ through the MTJ and the energy consumption $E$ is shown in Fig. 3(c) and (d), respectively, for both P-to-AP and AP-to-P switching for a unit probability with $t_{ON}$ equal to a respective $t_{Half}$. Because the $J$ is changing during the switching process, values are extracted when the $V_{MTJ}$ first reaches $V_0$ (see Fig. 1(b)). Consequently, the $J$ from P-to-AP is slightly larger than that from AP-to-P because of a different initial MTJ resistance. In addition, as expected, the $J$ increases linearly as the $V_0$ increases, and remains in the order of $10^8$ A-m$^{-2}$ because of a thick enough (2 nm) MgO layer. This low current density then ensures a relatively low switching energy as seen in Fig. 3(d). The switching energy has two components,

$$E = \int_0^{t_{ON}} \frac{(V_{MTJ}(t))^2}{R_{MTJ}(t)} dt + \frac{1}{2} \frac{\varepsilon_0 \varepsilon_{MgO} A}{t_{MgO}} V_0^2 , \qquad (10)$$

where $\varepsilon_{MgO}$ = 9.7 [70] is the relative permittivity of MgO, $\varepsilon_0$ is the vacuum permittivity, and *A* is the cross-sectional area of the MTJ. The first term in eq. (10) is the Joule heating $E_J$, and second term is the charging energy $E_C$ consumed by the MTJ capacitance. The capacitance has been assumed to be independent of the relative magnetization of PL and FL because of the sub-μm$^2$ cross-section of the MTJ [71]. $E_C$ ranges from 0.046 fJ to 1.138 fJ for $V_0$ from 0.6 V to 3 V, which is 9% to 14% of *E* (0.5 fJ to 8.1 fJ), respectively. Because of a large MTJ resistance at $t_{MgO}$ of 2 nm, the current density stays lower than $2\times10^9$ A-m$^{-2}$, and a low switching energy (<10 fJ/switch) is achieved (c.f. Table I of Ref. [69] for a general comparison with other non-volatile memory technologies). Moreover, Fig. 3(e) exhibits the switching probability from P-to-AP 'AND' AP-to-P as a function of $t_{ON}$ and $V_0$. Red regions denote a deterministic switching, i.e. 100% certainty of toggling. At $V_0$ = 1 V, the corresponding operation window of $t_{ON}$ is from 0.31 ns to 0.46 ns i.e. 0.15 ns, a decent margin for $t_{High}$ to vary without affecting the reliable operation. On the other hand, Fig. 3(f) presents the retention probability as a function of $t_{ON}$ and $V_0$. Red regions show with 100% probability that pre-configured data is not disturbed. As a consequence, the red regions can be used for reading, e.g. as long as the read voltage is below 0.4 V, the FL magnetic state will always remain unaltered, which implies that an absolutely disturb-free read operation can be achieved for the MRAM application.

## C. MgO Thickness Dependence

As demonstrated in the previous section, the device prefers a low-voltage operation for a decent pulse duration margin unto a value at which further decreasing $V_0$ may fail the switching. Hence, to study the $t_{MgO}$ dependence on the switching, $V_0$ is fixed to be 1.6 V in this section. This $V_0$ allows a sufficient $t_{MgO}$ variation range to achieve low energy consumption as expounded later. RA-products, TMRs and $V_{Half}$ for a different MgO thicknesses are extracted from the experimental paper [20]. For a fixed $V_0$ as the $t_{MgO}$ decreases, principally, both VCMA and STT effect become stronger. The former becomes stronger because the $E_z$ at the MgO-FL interface becomes stronger (see eq. (2)). Simultaneously, the RA-product decreases exponentially as the $t_{MgO}$ decreases. Since the current density $J$ is inversely proportional to the RA-product, $J$ thus increases exponentially from $10^7$ A-m$^{-2}$ to $10^{10}$ A-m$^{-2}$ for a decreasing $t_{MgO}$, as shown in Fig. 4(a), which makes the STT increasingly stronger. Next, the $t_{ON}$ in Fig. 4(b), which is chosen to equal $t_{Half}$, shows an increasing trend as $t_{MgO}$ increases because of a vitiating VCMA effect. At large $t_{MgO}$, the two curves overlap and a linear relation is observed. The linear relation is observed on account of the fact that $t_{ON}$ is inversely proportional to the precession frequency which in turn is almost proportional to $\mathbf{H_{Eff}}$. The $\mathbf{H_{Eff}}$ varies linearly with $\mathbf{H_K}$, and $\mathbf{H_K}$ is inversely proportional to $t_{MgO}$ (see eq. (2)) because of the VCMA effect. This indicates that the VCMA effect dominates over the STT effect for large $t_{MgO}$. For small $t_{MgO}$, there is a divergence in $t_{ON}$ between P-to-AP (black solid circles) and AP-to-P (red triangles) switching trends, implying that the STT effect is substantial which can be understood as follows. Beyond the critical electric field which changes the easy axis from z- to x-axis, further strengthening of the VCMA effect is inessential and has no significant additional contribution in the switching. However, STT effect has no such upper threshold in this case and starts to dictate the switching dynamics. For P-to-AP switching, since the electron flow direction is from FL to PL, due to the STT effect, the FL receives spin-flux anti-parallel to the magnetization of the PL. As a result, the STT effect assists the VCMA effect to attain an AP state for the FL and accelerates the switching process. In contrast, for AP-to-P switching, the STT effect contests with the VCMA effect to switch the FL. The FL still receives the spin-flux anti-parallel to the PL magnetization due to the STT effect and hence endeavors to maintain the FL in the AP-state. Conversely the VCMA effect strives to toggle the FL into a P-state. The two effects thus jostle to toggle the FL. This decelerates the toggling, which results in a larger $t_{ON}$ for the AP-to-P case. Moreover, because of the increasingly stronger dominance of the STT effect, the slope for the P-to-AP switching trend line increases as the $t_{MgO}$ decreases. However, for AP-to-P switching, on reducing the $t_{MgO}$, the $t_{ON}$ slope tends to transit from positive to negative.

As discussed in the previous section, the charging energy $E_C$ is a fraction of the Joule heating. When the STT effect is substantial, the net energy consumed (*E*) trend shown in Fig. 4(c) nearly follows the declining trend of the current density shown in Fig. 4(a). Nevertheless, when the $t_{MgO}$ becomes thicker than 2.2 nm, $E_C$ becomes comparable to $E_J$ because of the exponential decline in the current density and the corresponding Joule heating. Figure 4(c) also shows that as the STT effect wanes-off and VCMA dominates for a large $t_{MgO}$, the slope of *E* tapers down and *E* finally approaches 0.3 fJ, which is also the minimum energy achieved in this work. Figure 4(d) and (e), respectively, show the phase diagrams of the switching probability from P-to-AP (P10) and from AP-to-P (P01) as a function of $t_{ON}$ and $t_{MgO}$. An obvious oscillatory dependence on $t_{ON}$ is observed. At a large $t_{MgO}$ and for long pulse duration, the oscillations disappear because of the weak VCMA effect. Interestingly in Fig. 4(d), there is a sharp decrease in the $t_{ON}$ operation window for the first half precession cycle for a small $t_{MgO}$ (the red region on the left-bottom around $t_{MgO} = 1.6$ nm). This happens because in the said region a strong STT effect compliments the VCMA effect for P-to-AP switching, and greatly accelerates the switching process. This sharply reduces the precession period and the scope for tolerating variations in $t_{High}$. For a larger $t_{MgO}$, a wider operation window indicates that the device can tolerate more variations in $t_{High}$ in this regime. It can be found that it is more favorable to design the device with a large MgO thickness in the purview of a critical VCMA field for a given $V_0$, because both a larger margin in the pulse variation for the deterministic switching and a lower write-energy can be achieved in this regime.

**D. MTJ Scalability and Bias Magnetic Field**

As noted earlier, the external bias field required in the precession based VCMA switching has been a critical bottleneck in advancing it for memory applications and therefore in this work, the elliptical pMTJs have been presented so that $H_x$ can be engineered within the stack and provided by a BL. Besides the thickness of the BL discussed earlier, it is the cross-section of the MTJ which determines the number of magnetic moments in the BL for providing the bias-field through the FL that is to be toggled. Furthermore, the demagnetizing field scales with the cross-section thereby modifying the switching-time, required bias-field and energy landscape. Therefore, the MTJ cross-section is an important physical constraint to investigate to comprehend the device physics in the proposed VCMA device.

The effect of $H_x$ on the switching probability is shown in Fig. 5(a) and (b). For a pMTJ with FL of $150 \times 50 \times 1.7$ nm$^3$ dimensions, $t_{MgO}$ of 2 nm and $V_0$ of 1.6 V, there is a limited functional region in the range of 38-58 mT for $\mu_0 H_x$. This range can be shifted for different conditions. As seen from Fig. 5(a) and (b), increasing the bias field $H_x$ shrinks the red region, i.e. the operation window, which is similar to the case exhibited in Fig. 3(a) and (b). When the $\mu_0 H_x$ increases to more than 58 mT, an excessively strong **H$_{Eff}$** results in an overly fastened precession, thus sharply reducing the relaxation time, which is too fast to allow a deterministic switching. Conversely, if the $\mu_0 H_x$ is insufficient, i.e. less than 38 mT, a deterministic switching around the x-axis would not be supported. Hence, a probabilistic final state is attained by virtue of the thermal fluctuation. Approximately 50% switching probability is also observed when the pulse duration lasts long enough for a smaller $\mu_0 H_x$. This is because the large $t_{ON}$ provides enough time for the magnetization to relax along the x-axis. Once the voltage is removed, the easy-axis switches back to the z-axis, thus **m** would relax to either +z or −z-axis depending on the thermal fluctuation with equal probability. To reduce the required $H_x$ for switching, one possible way is to design the FL thickness even closer to the critical thickness but this would further sacrifice the thermal stability for instance for the NVL applications. This adjustment has three effects: it would weaken both the interfacial anisotropy and the demagnetizing field along the z-axis, and enhance the demagnetizing field along the x-axis. All of these would enable the operation at smaller $V_C$. In consequence, a stronger VCMA effect is obtained at the same $V_0$, which then relieves the requirement for a larger $\mu_0 H_x$.

Scalability of the pMTJ is next investigated in Fig. 5(c) and 5(d). The AR is held at 3, FL thickness at 1.7 nm, $t_{MgO}$ at 1.5 nm and $V_0$ at 1.6 V, while the length and width of the MTJ is swept. The bars in Fig. 5(c) represent the operation windows, which are extracted by simulating the devices 100 times under identical conditions (thermal fluctuations enabled). Within the operation window, switching happens with 100% certainty. Figure 5(c) shows that when the MTJ cross-section (represented as MTJ width) is scaled down, the optimal $t_{ON}$ (the data-markers on the curve), which equals respective $t_{Half}$, increases. This happens because on scaling down the MTJ, the $N_z$ decreases which thus increases the $K_{U\_Eff}$. As a result, the critical voltage $V_C$ (c.f. the inset), where $K_{U\_Eff} = 0$, as seen from eq. (2) and (8), becomes larger. In consequence, it is more difficult to switch. Hence, it takes larger $t_{ON}$ or the switching may even fail altogether. In addition, a considerable operation window is achieved for the MTJ cross-section between 39×117 nm$^2$ and 45×135 nm$^2$. On scaling down the MTJ cross-section below 39×117 nm$^2$, the deterministic switching fails because the switching becomes increasingly implausible, unless $V_0$ or $H_x$ is increased as discussed earlier. On the other hand, for the MTJ cross-section above ~ 45×135 nm$^2$, the operation window narrows as a result of the fastened precession process.

For the designs in Fig. 5(c), the respective energy consumption is shown in Fig. 5(d). At first a descending and then an ascending trend is observed. This behaviour is due to the competition between the $t_{ON}$ and the MTJ resistance as evinced in eq. (10). The former increases as observed in Fig. 5(c), while the latter also increases because for a given RA-product, the MTJ resistance increases as the MTJ cross-section is scaled down. These two have opposite contributions to the Joule heating; therefore, the trends exhibit a local minima where both $t_{ON}$ and MTJ resistance are optimized to deliver the lowest energy switching. These trends also imply that unduly scaling down the MTJ cross-section may not be attractive in terms of the energy consumption.

**IV. CONCLUSION**

We propose and appraise the elliptical pMTJs for a voltage controlled precessional switching. The $V_{MTJ}$, $t_{MgO}$, STT, bias magnetic field and MTJ scalability effects on the pMTJ properties are investigated. We show that an in-plane magnetized BL designed within the MTJ stack can bias the FL to eliminate the need of providing a uniform in-plane magnetic field for the FL by additional equipment or an external circuit. The pMTJ can be switched for as low as 0.3 fJ in just 0.38 ns at 1.6 V. Furthermore, it is shown that the Joule heating can be adequately suppressed by increasing $t_{MgO}$. We also find that the design favors to operate at low voltage (~ 1 V) and large MgO thickness. There is also a sufficient margin for the variation in $t_{High}$ without affecting the reliable operation. This should be encouraging for a practical disposition of the VCMA based MRAM. The advantages like fast switching, ultra-low energy consumption and non-volatility are very attractive for VCMA based MRAM application in the cache memories.

**Corresponding Authors**

*elelg@nus.edu.sg  †gauravdce07@gmail.com

**ACKNOWLEDGEMENTS**

This work at the National University of Singapore was supported by CRP award no. NRF-CRP12-2013-01 and MOE2013-T2-2-125. We gratefully acknowledge the discussions with Xuanyao Fong, and the funding support from the National Research Foundation, Prime Minister Office, Singapore, under its Medium Sized Centre Programme.

**Figures**

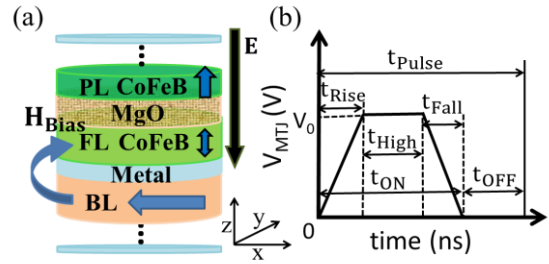

**Figure 1**. (a) Schematic of an elliptical pMTJ with the major axis along the x-axis. A bias layer (BL) separated by a metal is engineered to provide the exchange field for the FL. The top and bottom water-blue layers are electrodes. (b) The applied voltage pulse with $t_{Rise} = t_{Fall}$ = 50 ps and $t_{OFF}$ of 10 ns, which is long enough to ensure that the magnetization is completely relaxed. Positive voltage is defined for the electric field pointing from the PL to the FL.

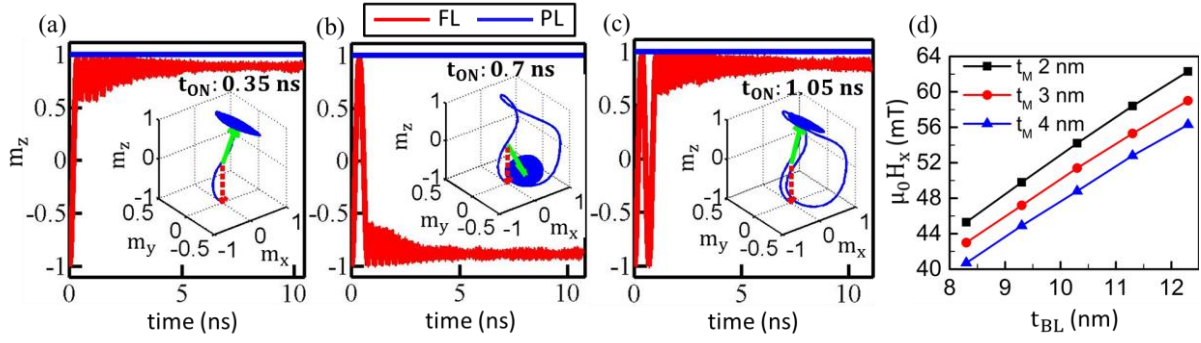

**Figure 2**. FL switching for a voltage pulse with $t_{ON}$ in integer multiples of the half precession periods ($t_{Half}$): $t_{ON}$ is (a) 0.35 ns, (b) 0.7 ns and (c) 1.05 ns while $t_{Half}$ is 0.35 ns. For the odd multiples the FL toggles while for the even cases it restores to the original state and remains unaltered. Insets show a full three dimensional (3D) FL dynamics for the respective cases, where the red dash arrow, green arrow and blue line is the initial **m** state, final **m** state and its trajectory, respectively. (d) The exchange field acting on the FL along the x-axis generated by the BL for different metal spacer thicknesses $t_M$ and BL thicknesses $t_{BL}$. This large enough exchange field, e.g. 50 mT for the $t_{BL}$ of 10 nm and the $t_M$ of 3 nm, can bias the FL for the precession based VCMA switching.

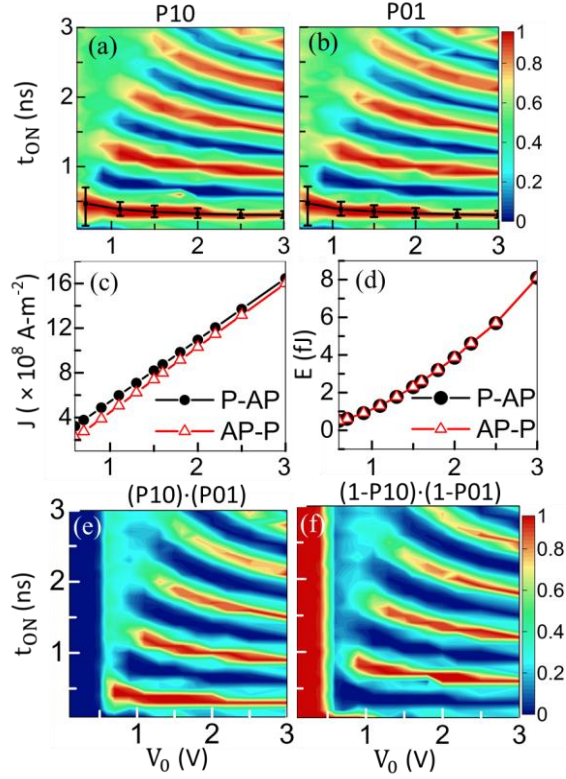

**Figure 3**. Phase diagram for switching probability from (a) parallel (P) to anti-parallel (AP) state (P10) and from (b) AP-to-P (P01) as a function of $t_{ON}$ and $V_0$ for a $t_{MgO}$ of 2 nm and a bias field $\mu_0 H_x$ of 50 mT (with thermal fluctuation). The black bars show the $t_{ON}$ operation windows without considering the thermal fluctuation. The increasing operation window as the $V_0$ decreases indicates that the device favors a low-voltage operation. (c) Current density and (d) energy consumption per switch *vs*. $V_0$. (e) Switching probability from P-to-AP 'AND' AP-to-P (P10·P01) as a function of $t_{ON}$ and $V_0$. Red regions indicate deterministic switching, i.e. 100% certainty of toggling. (f) Retention probability, i.e. (1-P10)·(1-P01) as a function of $t_{ON}$ and $V_0$. Red regions show with 100% probability that pre-configured data would not be altered.

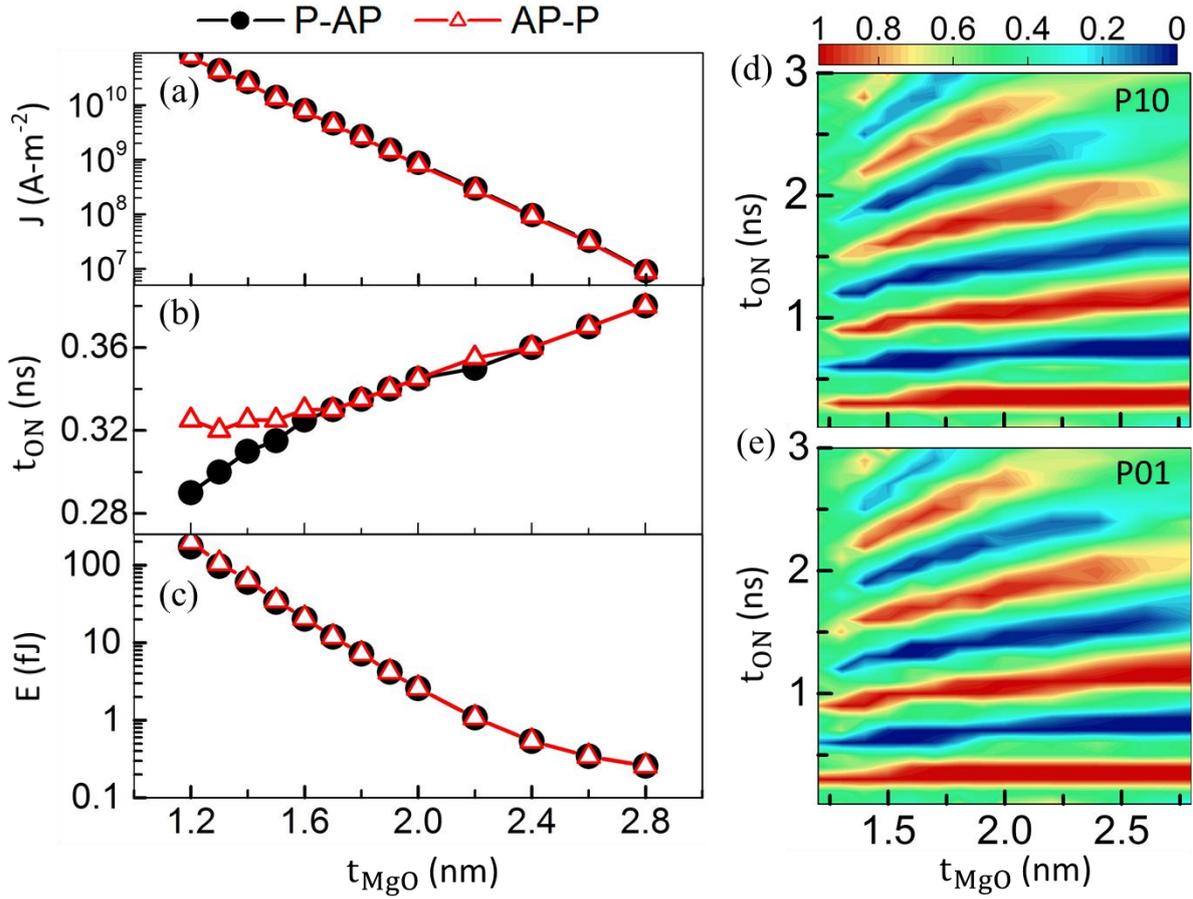

**Figure 4**. Effects of $t_{MgO}$ on (a) current density, (b) $t_{ON}$ (optimal values used to switch, i.e. $t_{Half}$) and (c) energy consumption per switching operation at 1.6 V $V_0$ and 50 mT $\mu_0 H_x$. Red lines with open triangles show the process from AP-to-P. Black lines with solid circles correspond to the process from P-to-AP. The smallest switching energy achieved is 0.3 fJ. Phase diagram for switching probability from (d) P-to-AP and (e) AP-to-P *vs.* $t_{ON}$ and $t_{MgO}$.

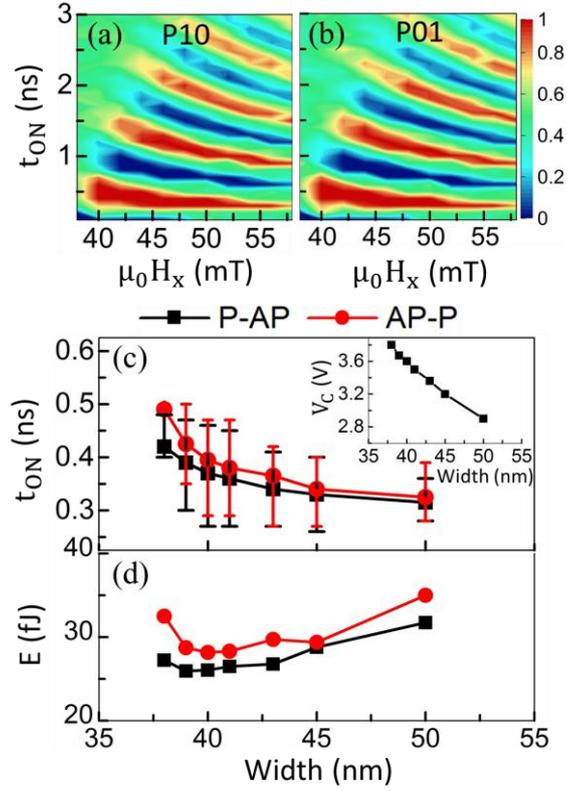

**Figure 5**. Phase diagram for switching probability of (a) P-to-AP and (b) AP-to-P as a function of $t_{ON}$ and bias magnetic field $\mu_0H_x$. $H_x$ is the magnitude of $\mathbf{H}_{Bias}$ projecting along x-axis. Optimal $t_{ON}$ used to switch ($t_{Half}$) in (c) and energy consumption in (d) as a function of the MTJ width. The AR and FL thickness are held constant at 3 and 1.7 nm respectively, e.g. for width of 40 nm, the MTJ cross-section is 40×120 nm$^2$. The inset shows the critical voltage $V_C$ *vs*. the MTJ width.